\documentclass[conference]{IEEEtran}
\IEEEoverridecommandlockouts
\usepackage{cite}
\usepackage{amsmath,amssymb,amsfonts}
\usepackage{algorithmic}
\usepackage{hyperref}
\hypersetup{
    colorlinks=true,
    linkcolor=blue,
    filecolor=black,      
    urlcolor=blue,
}
\usepackage{arydshln}
\usepackage{graphicx}
\usepackage{textcomp}
\usepackage{xcolor}
\usepackage{multirow}
\usepackage{booktabs}
\usepackage{epstopdf}

\def\BibTeX{{\rm B\kern-.05em{\sc i\kern-.025em b}\kern-.08em
    T\kern-.1667em\lower.7ex\hbox{E}\kern-.125emX}}
\begin{document}

\title{Hierarchical Clustering of World Cuisines}

\author{
\IEEEauthorblockN{1\textsuperscript{st} Tript Sharma}
\IEEEauthorblockA{\textit{Department of Mechanical Engineering} \\
\textit{Delhi Technological University}\\
New Delhi, India \\
triptsharma22@gmail.com}
\and
\IEEEauthorblockN{1\textsuperscript{st} Utkarsh Upadhyay}
\IEEEauthorblockA{\textit{Department of Electronics Engineering} \\
\textit{Jamia Millia Islamia University}\\
New Delhi, India \\
utkarshdhy@gmail.com}
\and
\IEEEauthorblockN{2\textsuperscript{nd} Jushaan Kalra}
\IEEEauthorblockA{\textit{Department of Computer Engineering} \\
\textit{Delhi Technological University}\\
New Delhi, India\\
jushaan18@gmail.com}
\and
\IEEEauthorblockN{3\textsuperscript{rd} Sakshi Arora}
\IEEEauthorblockA{\textit{Department of Computer Science} \\
\textit{Indraprastha Institute of Information and Technology }\\
New Delhi (IIIT-Delhi), India \\
sakshi18133@iiitd.ac.in}\
\and
\IEEEauthorblockN{4\textsuperscript{th} Saad Ahmad}
\IEEEauthorblockA{\textit{Department of Computational Biology} \\
\textit{Indraprastha Institute of Information and Technology }\\
New Delhi (IIIT-Delhi), India \\
saad18409@iiitd.ac.in}\
\and
\IEEEauthorblockN{4\textsuperscript{th} Bhavay Aggarwal}
\IEEEauthorblockA{\textit{Department of Computational Biology} \\
\textit{Indraprastha Institute of Information and Technology }\\
New Delhi (IIIT-Delhi), India \\
bhavay18384@iiitd.ac.in}\
\and
\and
\IEEEauthorblockN{\textsuperscript{} Ganesh Bagler}
\IEEEauthorblockA{\textit{Center for Computational Biology} \\
\textit{Indraprastha Institute of Information and Technology}\\
New Delhi (IIIT-Delhi), India \\
bagler@iiitd.ac.in}
}
\maketitle

\begin{abstract}
Cultures across the world have evolved to have unique patterns despite shared ingredients and cooking techniques. Using data obtained from RecipeDB, an online resource for recipes, we extract patterns in 26 world cuisines and further probe for their inter-relatedness. By application of frequent itemset mining and ingredient authenticity we characterize the quintessential patterns in the cuisines and build a hierarchical tree of the world cuisines. This tree provides interesting insights into the evolution of cuisines and their geographical as well as historical relatedness. 
\end{abstract}


\section{Introduction}
Cultures across the world have evolved diverse cooking practices over time. Although the underlying fundamentals of cooking remain the same, various factors including geography and climate have affected cooking styles. Cuisines from across the globe have thus acquired their signature styles. Each cuisine has interesting patterns that are inherent to it while sharing some common attributes with others. In this article, we characterize the unique features that typify every cuisine in an attempt to  discern the footprint of food on human cultures and inter-relatedness of world cuisines. 


\section{Literature Survey}
With increasing availability of recipes data, there has been much interest in data mining recipes data. One of the focus has been on defining recipe similarity. Attempts have been made to define similarity based on various elements of cooking recipes~\cite{x} and ingredients~\cite{y,z}.

Among other efforts in data mining recipes data have focused on food pairing phenomena in cuisines. Among one of the early studies, Shidochi et al~\cite{new1} experimented with the possible replacements of ingredients in a recipe. Jain et al~\cite{jain2015analysis} investigated the phenomenon of food pairing which examines compatibility of two ingredients in a recipe in terms of their shared flavor compounds. This study investigated the food pairing phenomena in Indian recipes and proclaimed that spices form the basis of their food pairing. The work was extended by Singh et al~\cite{singh2018data} to analyze a much larger dataset encompassing 22 cuisines across the world and found interesting food pairing patterns in cuisines from across the world. Tuwani et al \cite{tuwani2019computational}, on the other hand, considered culinary systems as a function of socio-cultural factors and presented computational models for cuisine evolution. An interesting work by Yokoi et al~\cite{new2} calculated an ingredient associative metric called `typicality value', giving out typical recurring ingredient patterns.

In this article, we indulge in pattern analysis in world-wide cuisines by way of association rule discovery and frequent pattern mining \cite{marketbasketa}. Going beyond the application of pattern mining techniques on cuisines in \cite{ozaki2017extraction}, we propose their use for frequent pattern mining of recipe data and cooking processes, utensils and ingredients for hierarchical clustering of cuisines.

\section{Data Collection}
\label{data}
Our analysis involved four types of information pertaining to traditional recipes, namely, recipes, ingredients, processes and utensils. A total of 118,071 recipes were obtained from various sources:  \href{https://www.allrecipes.com}{AllRecipes}, \href{https://www.foodnetwork.com}{Food Network}, \href{https://www.epicurious.com}{Epicurious} and \href{https://www.tarladalal.com} {TarlaDalal}. RecipeDB \cite{recipedb}, a structured compilation of recipes, was used as the primary source of information. All data are available at \href{https://cosylab.iiitd.edu.in/recipedb}{`RecipeDB: A resource for exploring recipes'}.

\newcommand\Tstrut{\rule{0pt}{2.6ex}}
\def\arraystretch{1.1}
\begin{table*}[htbp]
  \centering
  \caption{Significant Patterns mined from cuisines across the world}
    \begin{tabular}{|l|c|p{12.93em}|c|c|}
    \hline
    Region & Number of Recipes & Pattern & Support & \multicolumn{1}{l|}{Number of patterns}\Tstrut \\
\hline

    Australian & 5,823  & Butter & 0.24 & 29 \\

    Belgian & 1,060  & Butter + salt & 0.24 & 51 \\

    Canadian & 6,700  & Onion & 0.20 & 31 \\
    Caribbean & 3,026  & Garlic Clove & 0.24 & 32 \\
    Central American & 460   & Onion &
    0.30 & 38 \\
    Chinese and Mongolian & 5,896  & Soy sauce + add + heat & 0.27 & 88 \\
    Deutschland & 4,323  & Onion & 0.29 & 54 \\
    Eastern European & 2,503  & Cream & 0.30 & 60 \\
    French & 6,381  & skillet & 0.21 & 60 \\
    Greek & 4,185  & Olive Oil & 0.40& 43 \\
    Indian Subcontinent & 6,464  & Onion + add + heat + salt &0.22 & 119 \\
    Irish & 2,532  & Butter & 0.32 & 41 \\
    Italian & 16,582 & Parmesan cheese & 0.31 & 63 \\
    Japanese & 2,041  & Soy Sauce & 0.45 & 45 \\
    Mexican & 14,463 & cilantro & 0.25 & 33 \\
    Rest Africa & 2,740  & Onion + add + heat & 0.20 & 51 \\
    South American & 7,176  & Onion + salt & 0.21 & 62 \\
    
    Southeast Asian & 1,940  & Fish sauce & 0.24 & 69 \\
    
    Spanish and Portuguese & 2,844  & Olive Oil & 0.31 & 67 \\
    
    Thai  & 2,605  & Fish sauce + add + heat & 0.23 & 73 \\
    \hdashline
    \multirow{2}{*}{Korean} & \multirow{2}{*}{668} & Soy sauce + sesame oil  & 0.34 & \multirow{2}{*}{85}\\
    & & green onion + sesame oil & 0.24&   \\
    \hdashline

    \multirow{2}{*}{Middle Eastern} & \multirow{2}{*}{3,905} & Salt + bowl  & 0.22 & \multirow{2}{*}{46}\\
    & & Lemon Juice & 0.22&   \\
    \hdashline
    
    \multirow{3}{*}{Northern Africa} & \multirow{3}{*}{1,611} & cumin + cinnamon  & 0.21 & \multirow{3}{*}{134}\\
    & & cumin + olive oil & 0.22&   \\
    & & cumin + Salt & 0.22 & \\
   
    \hdashline
    \multirow{2}{*}{Scandinavian} & \multirow{2}{*}{2,811} & Butter + Salt  & 0.22 & \multirow{2}{*}{52}\\
    & & Salt + Sugar & 0.21&   \\

    \hdashline
    \multirow{2}{*}{UK} & \multirow{2}{*}{4,401} & Butter  & 0.37 & \multirow{2}{*}{45}\\
    & & Oven & 0.46&   \\
    \hdashline
    \multirow{2}{*}{US} & \multirow{2}{*}{5,031} & Bake + preheat+ oven + bowl  & 0.22 & \multirow{2}{*}{67}\\
    & & Onion & 0.25&   \\
    \hline
    \bottomrule
    \end{tabular}%
  \label{tab:patterns}%
\end{table*}%

For each recipe, details such as its name and the list of ingredients and processes involved while cooking are available. Each recipe was treated as an unordered list of ingredients, processes and utensils. We integrated recipes from all the sources and grouped them into 26 distinct geo-cultural `cuisines' while ensuring that each region had enough recipes attributed to it to distinguish it as a cuisine. Please refer to Table~\ref{tab:patterns} for the list of regions. Due to insufficient information about the region for many recipes, they were aggregated together on the basis of their geological similarities with the prefix `Rest'. For example, recipes without `region' information belonging to Africa were put in `Rest Africa' category.

The database consists of 20,280 unique ingredients, 268 unique processes and 69 unique utensils. The data are sparse in the list of utensils and 14,601 recipes don't have information regarding the preferred utensils required for cooking. An average recipe in a cuisine has $\sim$10 ingredients, $\sim$12 processes and $\sim$3 utensils. This is intuitive as too many ingredients would impede the success/propagation of a recipe, whereas too few would lead to it being modified easily. Thus recipes needs to maintain a balance between complexity and simplicity to survive successive iterations of evolution \cite{tuwani2019computational}. 

\section{Pattern Mining}
To investigate the ontology of food, we mined rules from the data to understand the patterns that are prevalent in a given cuisine. The methodology employed for mining patterns is explained in Section~\ref{method}. The mined patterns consist of ingredients, processes and utensils permutations that have a frequency greater than the defined threshold support. According to~\cite{fpgrowth, marketbasketa}, support represents the frequency with which the collection of items co-occur as a percentage of all transactions. A high support threshold represents high confidence in the pattern being mined, whereas with a low support threshold noise can creep into the mined patterns, leading to false identification of cuisine features. Hence, a trade off support of 20\% was chosen as the threshold. Since the mined patterns are the most frequent ones, it is safe to say that most of the recipes follow the observed  patterns and essentially define the cooking practices of the cuisine.

Among the patterns obtained in all the recipes for the Korean region as shown in Table \ref{tab:patterns}, the pattern ``Soy Sauce + sesame oil'' occurs with a support of 0.34, i.e. the pattern is found in 34\% of all the recipes in the Korean region. Table~\ref{tab:patterns} contains the topmost significant patterns in the 26 cuisines.
The pattern depicts set of words occurring in a particular recipe. The patterns mined are highly skewed, with most regions containing patterns having generic ingredients such as `salt', `onion' and processes such as `add' and `cook', which is justified as they have a high frequency among all cuisines and are fundamental to cooking in many cuisines.

\section{Methodology}
\label{method}
We implemented two approaches namely Frequent Itemset-based Hierarchical Clustering (FIHC)~\cite{fung2003hierarchical} and Authenticity based Clustering~\cite{auth} to extract relationships between various cuisines of the world. The hypothesis is that some patterns which are common across a subset of cuisines would be found, which defines their `similarity/closeness' with each other.

\subsection{Frequent Pattern Mining}
Frequent Itemset Mining refers to discovering interesting patterns in databases such as association rules from a set. Since we treat a recipe as a combination of ingredients, processes and utensils, it can be treated as an unordered set of these entities. For the frequent itemset mining, the FP-Growth Algorithm~\cite{fpgrowth} was used as it is an efficient and scalable method for mining the complete set of frequent patterns by pattern fragment growth. 
The data extracted from RecipeDB was pre-processed to make it compatible with the input form of FP-Growth Algorithm. Ingredients, utensils and processes were concatenated and the FP-Growth Algorithm was applied.
This approach was applied to all 26 regions present in the data extracted from RecipeDB. The support was kept at 0.2 so that the pattern was mined across a reasonable number of recipes. 

\subsection{Authenticity Based Clustering}
We propose that a cuisine can be represented as a set of ingredients, process and utensils which can thus be utilized to define the relationships among the cuisines. Using the authenticity metric described in~\cite{auth} we calculate the prevalence $P_i^c$ of an item $i$ in a cuisine $c$ according to equations~\ref{auth1} as a function of number of recipes, $n_i^c$ in a cuisine over total number of recipes in the dataset, $N_C$. This is used to calculate the authenticity of the item for a cuisine using equation~\ref{auth2}.

\begin{equation}
    P_i^c = n_i^c/N_C
    \label{auth1}
\end{equation}
\begin{equation}
    p_i^c = P_i^c - (P_i^k)_{c \neq k}
    \label{auth2}
\end{equation}

In order to obtain the contribution of the item in uniquely identifying a cuisine, a relative prevalence matrix is created by subtracting the average prevalence of the item, say $i$ for all cuisines from the prevalence for cuisine $c$. Accordingly, the most prevalent and least prevalent items in a cuisine can be identified. It should be noted that both the most prevalent and least prevalent items would contribute towards the culinary fingerprint of a cuisine as the former indicates the items having a relatively higher utility in the cuisine while the latter indicates items that are least used in the cuisine versus the rest of the world cuisines.

\section{Clustering Techniques}

\subsection{Hierarchical Clustering}
Application of FP-Growth Algorithm on the prepared dataset results in 26 files, each containing patterns in a `frozenset' along with their respective support to remove redundant patterns. These patterns were extracted from the `frozenset' and appended together in a list in a sorted fashion. All the elements of this list are appended and converted into a string resulting in a `string pattern'. All the `string patterns' are compiled into a set resulting in unique set of patterns across all the 26 regions. Since the data is in string form and each element is a unique entity, it can be classified as a category. Therefore, unique set of `string patterns' are fit for using Label Encoding (because the strings are categorical data) to get a transformer and the `string patterns' in the rules are transformed using the derived transformer across all the regions. All the `string patterns' in the rules from all the regions are appended in an array. The data from this array is thus vectorized to form a feature vector which is thus fed to the cluster as the linkage matrix.

Three different approaches were applied in order to cluster the linkage matrix data and to generate subsequent dendrograms. The linkage matrix is converted into a condensed distance matrix (pdist) in order to calculate the distance between all the cuisines based on the rules mined and is then fed into the hierarchical clustering model. To analyze the clusters we have used three distance metrics:

\begin{equation}
    \text{Jaccard Distance} = \frac{c_i \cup c_j}{c_i \cap c_j}
\end{equation}
 
\begin{equation}
    \text{Cosine  Distance} = \frac{c_i.c_j}{|c_i||c_j|}
\end{equation}
  
\begin{equation}
    \text{Euclidean Distance} = \sqrt{c_i^2 + c_j^2}
\end{equation}
\noindent
where cuisines $c_i$, $c_j$ $\in$ $C$, the universal set of cuisines in the dataset. To calculate the distance between two cuisines, they must be quantified. This was done by vectorizing the patterns obtained by the above-mentioned pre-processing technique.

\begin{figure}[bp]
    \centering
    \hspace*{-0.15in}
    \includegraphics[scale=.4]{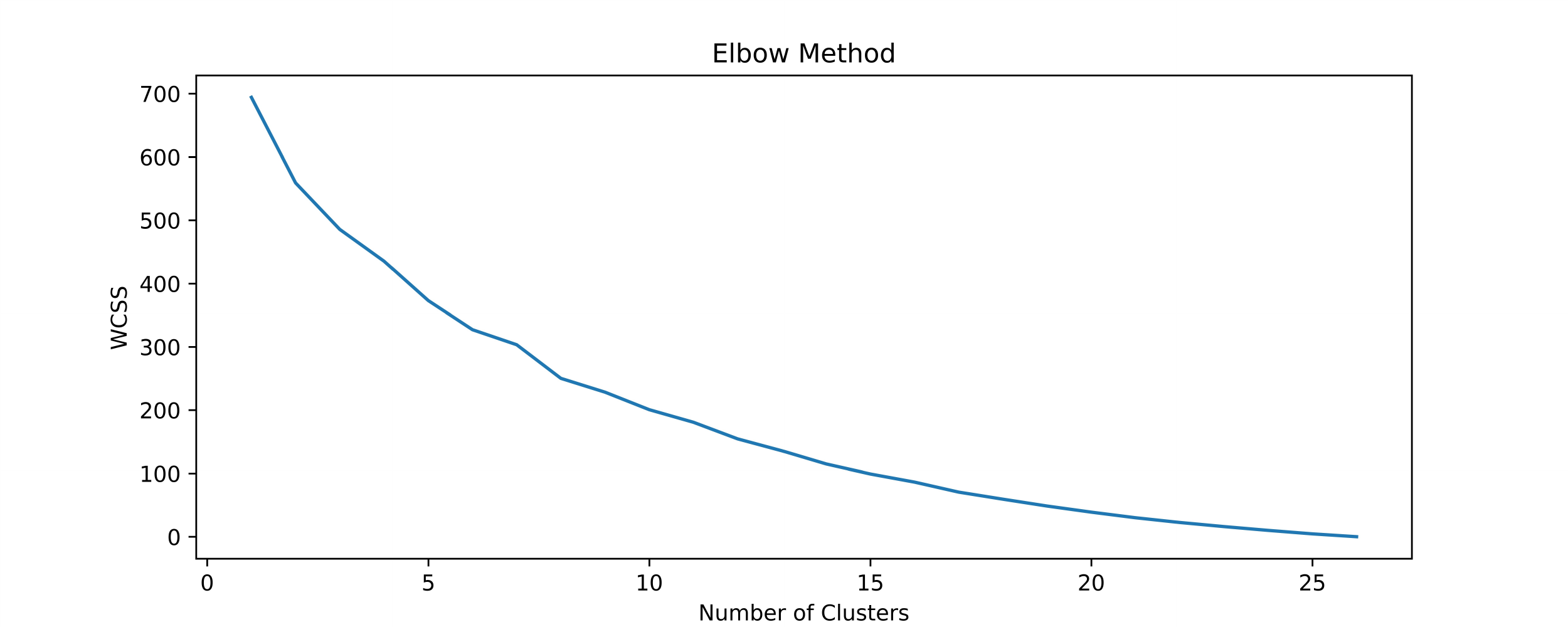}
    \caption{Elbow Method for cluster identification}
    \label{fig:kmeans}
\end{figure}

\begin{figure*}[htbp]
    \centering
    \hspace*{-.2in}
    \includegraphics[scale=.6]{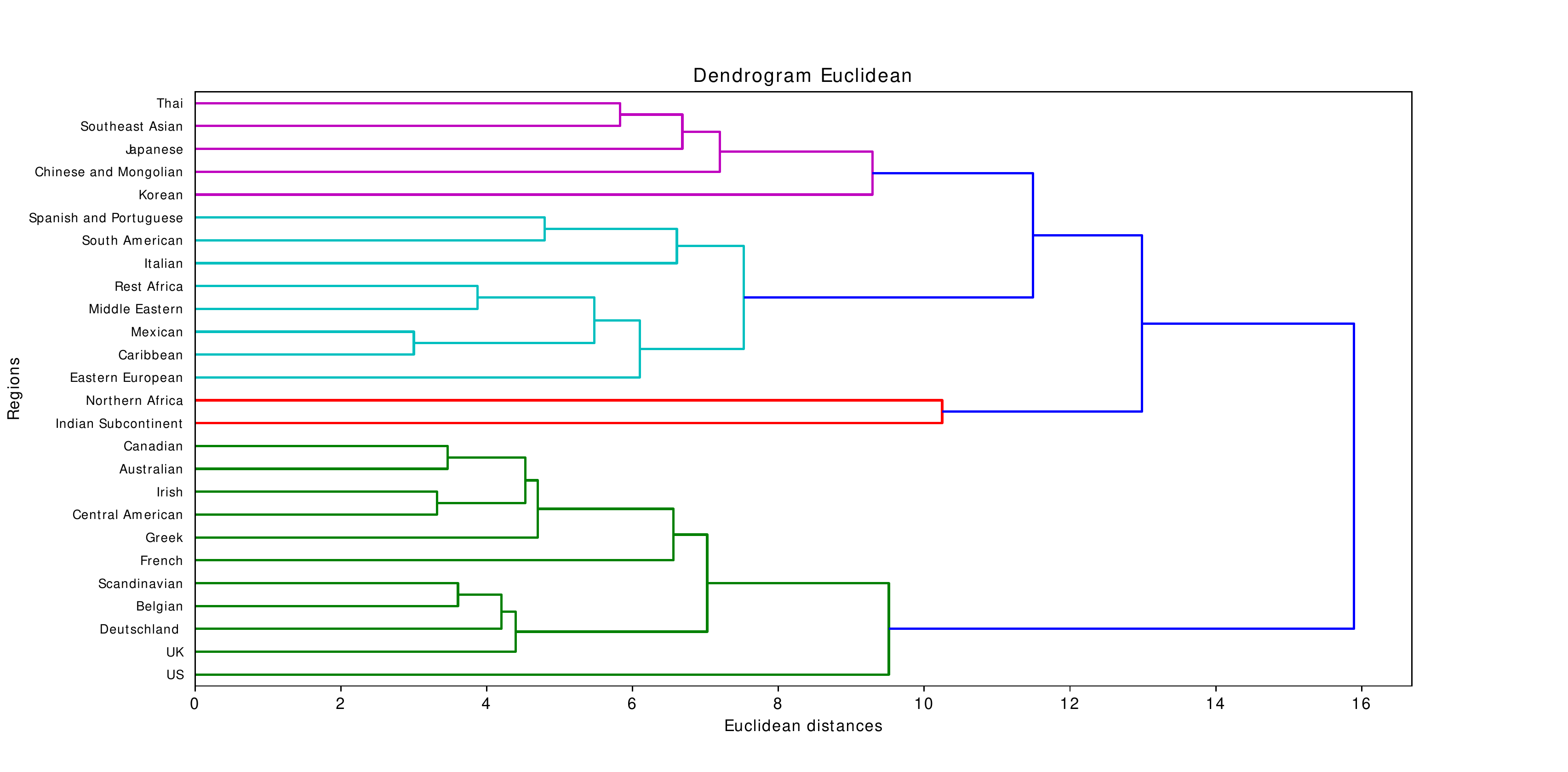}
    \caption{Hierarchical Agglomerative Clustering based on Patterns Mined using Euclidean distance
    }
    \label{fig:euc}
\end{figure*}


\begin{figure*}[htbp]
\minipage{.45\textwidth}
  \hspace*{-2.2in}
    \includegraphics[height=300pt]{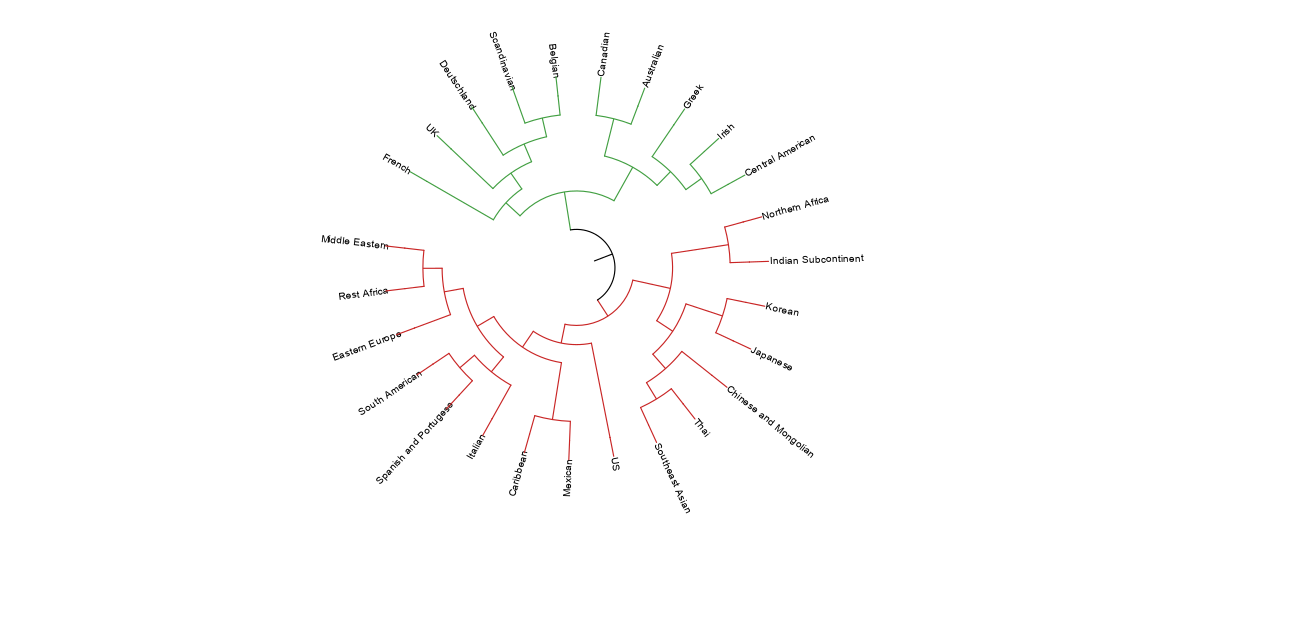}
  \caption{Hierarchical Agglomerative Clustering based on Patterns Mined using Cosine distance}\label{fig:cos}
\endminipage\hfill
\minipage{.45\textwidth}
  \hspace*{-1.7in}
  \includegraphics[height=300pt]{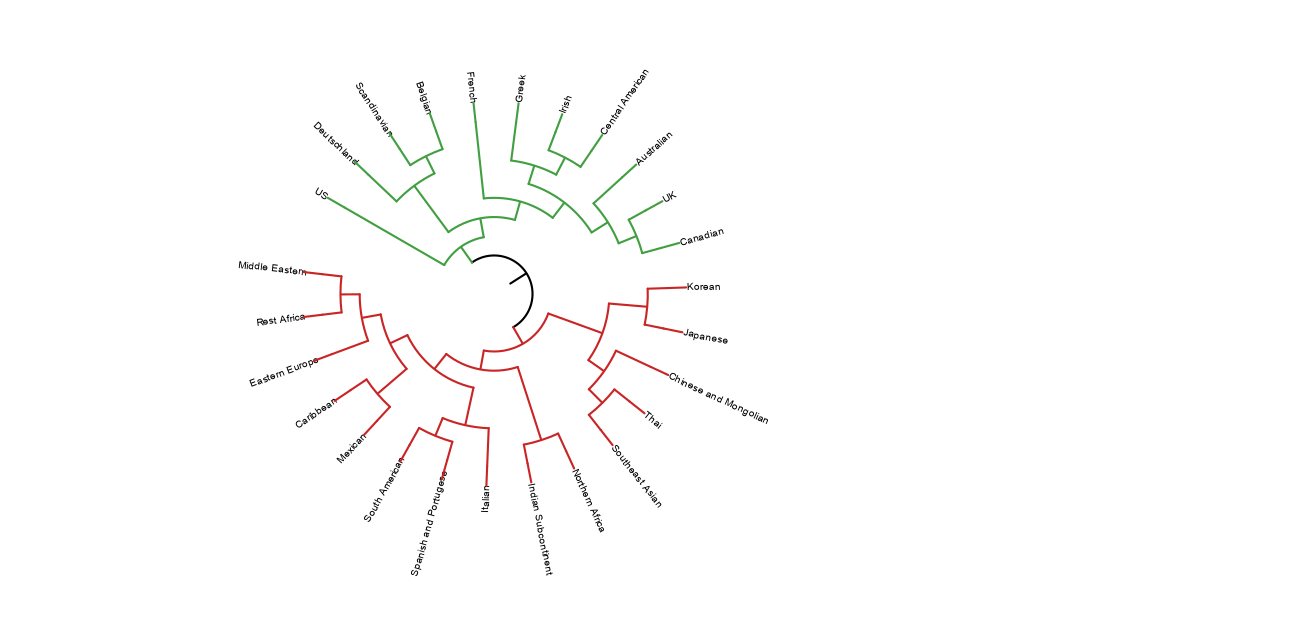}
  \caption{Hierarchical Agglomerative Clustering based on Patterns Mined using Jaccard distance}\label{fig:jac}
\endminipage
\end{figure*}

\begin{figure*}[htbp]
    \centering
    \includegraphics[scale=.5]{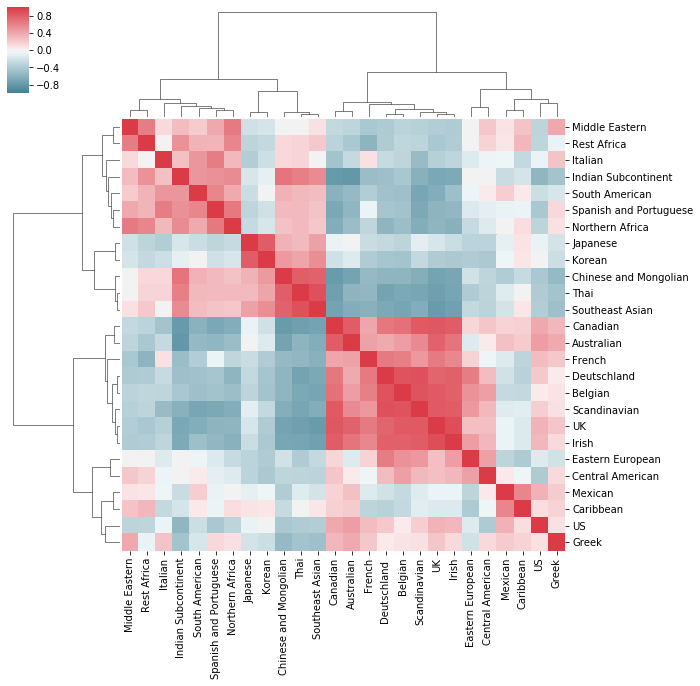}
    \caption{Hierarchical Agglomerative Clustering based on Authenticity of Ingredients}
    \label{fig:auth}
\end{figure*}

\begin{figure*}[htbp]
      \centering
      \hspace*{-1.5in}
      \includegraphics[scale=.5]{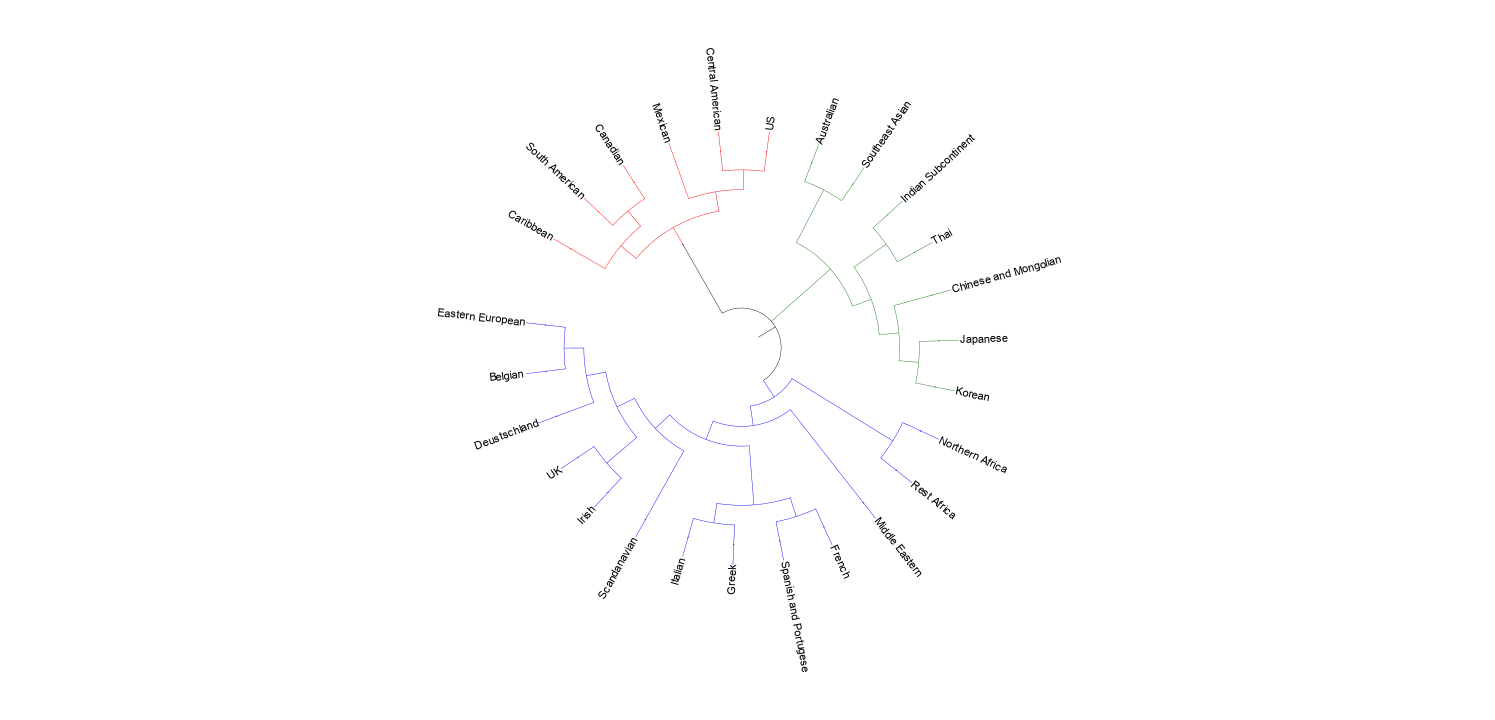}
      \caption{Hierarchical Agglomerative Clustering based on Geographical Distance of Regions}
      \label{geo}
\end{figure*} 

\subsection{K-means Clustering}
Another popular clustering technique, K-means, was applied on our categorical data. It has been shown in \cite{hac_kmean} that hierarchical agglomerative  clustering is a better approach for clustering categorical data than K-means. The elbow analysis and the subsequent WCSS score on our dataset indicates similar results. The elbow method~\cite{elbow} analysis fails to determine the number of appropriate clusters for our dataset. As in Figure \ref{fig:kmeans}, no sharp edge or elbow like structure is obtained which determines the number of clusters. While on the other hand the  hierarchical agglomerative clustering technique presents with better cluster representation. Therefore, our results were predominantly determined by hierarchical agglomerative clustering.

\section{Results}
To evaluate the efficiency of the proposed methodologies, the RecipeDB dataset mentioned in Section~\ref{data} was used. This dataset was used to identify the patterns which were then fed into the Sequential Pattern Mining based clustering algorithm while the ingredients obtained from the dataset were the input features for the Authenticity-based clustering. The corresponding code and relevant files are present in the \href{https://github.com/cosylabiiit/Hierarchical-Clustering-Ingredients}{GitHub repository (https://github.com/cosylabiiit/Hierarchical-Clustering-Ingredients)}.

The Hierarchical Agglomerative clustering (HAC) gives clusters for all regions based on the three approaches and presents a cluster dendrogram for each approach. Figures~\ref{fig:euc}, \ref{fig:cos} and \ref{fig:jac} represent the clusters formed using the feature vector obtained via Euclidean, Cosine and Jaccard metrics for pairwise distance calculation respectively. Similarly, Figure~\ref{fig:auth} shows the authenticity based approach to determine the correlation of cuisine and regions, dominantly based on ingredients.

Because of the absence of a quantified validation metric for cuisine similarity, the geographical relationship among the cuisines was used to validate the accuracy in the prediction of cuisine interrelationships. It is observed that while comparing the Figures \ref{fig:euc}, \ref{fig:cos} and \ref{fig:jac} with Figure \ref{geo} the results received from the Euclidean distance model were most similar to the geographical distribution of the countries. On the other hand, the clusters obtained via the authenticity based clustering gave similar yet better results than Euclidean distance-based HAC when validated on geographical distance based clusters.

Authenticity-based Clustering identifies both positive and negative relationships between cuisines and items whereas pattern based techniques take only the positive relationships into account. This leads to the difference in the results. Yet, despite the differences, both techniques predict a closer relationship among Canadian and French cuisines as compared to Canadian and US cuisines despite their geographical proximity. This is evident from the historical fact that Canada was a French colony. Another interesting grouping is that of Indian Subcontinent and Northern Africa. Due to prevalent use of spices in the two regions, Indian subcontinent cuisine is closer to African cuisine as compared to its geographical neighbors like Thai and Southeast Asian cuisines. Hence, the obtained clusters are also able to identify relationships deviating from the geographical similarities.

\section{Conclusions and Future Work}
In this exploratory work we proposed and analyzed two methodologies for fingerprinting cuisines and identifying their interdependence. Our clustering algorithms show how various cuisines are interrelated and show trends similar to their geographical associations. It shows how cooking practices and methods are shared by neighbouring regions. This analysis is important from a historical and cultural point of view as it helps in appreciating how cooking practices are distributed across the world. Furthermore, we also provide a verbose list of patterns identified in the cuisines. These patterns include compound patterns; combination of ingredients, processes and utensils that can be used to identify the relationship among these items.

While this article introduces new methods for investigation of cuisine correlations, it raises new research questions. How do factors such as climate, economy and genetics influence the cuisine patterns? RecipeDB is a sparse dataset in terms of utensils and processes. Hence, to what extent do they influence the relationships among cuisines is yet to  be answered. Among one of the limitations of this study, it neither considers the state of ingredients nor their aliases. Therefore, future analysis need to account for the aliases along with state of ingredients and other properties like cooking time and preparation time for the task. It would also be interesting to identify more sophisticated validation metric for cuisine ontology than geographical clustering.

We believe that this study can be applied for cuisine fingerprinting, food ontology and exploration of relations between food and culture. Probing the past and present interrelatedness among cuisines can provide insight into human behavior and cultures, and means for shaping the future of food. 

\section{Acknowledgement}
G.B. thanks the Indraprastha Institute of Information
Technology (IIIT-Delhi) for providing computational facilities and support. T.S, U.U, J.K,S.A, S.A. and B.A. are Research Interns in Dr. Bagler's lab (Complex Systems Laboratory) at the Center for Computational Biology. All the research interns are thankful to IIIT-Delhi for the support.

\bibliographystyle{IEEEtranS}

\bibliography{biblo}

\end{document}